\documentclass{mem}
\usepackage{natbib}\usepackage{txfonts}\usepackage{balance}
\usepackage{graphicx}
\usepackage[a4paper,breaklinks,dvipdfm]{hyperref}
\idline{75}{1}
\begin{document}
\newcommand{\Teff}{T_{\rm eff}}
\newcommand{\logg}{\rm log~ g}
\def\kms{$\mathrm {km s}^{-1}$}
\newcommand{\Vmac}{V_{\rm mac}}
\newcommand{\Eexc}{$E_{\rm exc}$}
\newcommand{\kH}{$S_{\!\!\rm H}$}    
\newcommand{\avr}{$\ensuremath{\langle\mathrm{3D}\rangle}$}
\newcommand{\eps}[1]{\log\varepsilon_{\rm #1}}

\title{Signs of atmospheric inhomogeneities \\ in cool stars \\ from 1D-NLTE analysis of iron lines
}

   \subtitle{}


\author{
L.\,Mashonkina\inst{1}
\and H.-G.\,Ludwig\inst{2}
\and A.\,Korn\inst{3}
\and T.\,Sitnova\inst{1}
\and E.\,Caffau\inst{2}
}

  \offprints{L. Mashonkina}

\institute{
Institute of Astronomy, Russian Academy of Sciences, RU-119017 Moscow,
     Russia \\ \email{lima@inasan.ru}
\and Zentrum f\"ur Astronomie der Universit\"at Heidelberg, Landessternwarte,
     K\"onigstuhl 12, D-69117 Heidelberg, Germany
\and Department of  Physics and Astronomy, Division of Astronomy and Space Physics, Uppsala University, Box 516, 75120 Uppsala, Sweden
}

\authorrunning{Mashonkina}

\titlerunning{Signs of atmospheric inhomogeneities in metal-poor stars}

\abstract{For the well studied halo star HD\,122563 and the four stars in the globular cluster NGC\,6397, we determine NLTE abundances of iron using classical plane-parallel model atmospheres. Each star reveals a discrepancy in abundances between the \ion{Fe}{i} lines arising from the ground state and the other \ion{Fe}{i} lines, in qualitative agreement with the 3D-LTE line formation predictions, however, the magnitude of the observed effect is a factor of 2 smaller compared with the predicted one. When ignoring the \ion{Fe}{i} low-excitation lines, the NLTE abundances from the two ionization stages, \ion{Fe}{i} and \ion{Fe}{ii}, are consistent in each investigated star. For the subgiants in NGC\,6397, this is only true when using the cooler effective temperature scale of \citet{Alonso1999}.
We also present full 3D-LTE line formation calculations for some selected iron lines in the solar and metal-poor 4480/2/$-3$ models and NLTE calculations with the corresponding spatial and temporal average \avr\ models. The use of the \avr\ models is justified only for particular \ion{Fe}{i} lines in particular physical conditions. Our NLTE calculations reproduce well the centre-to-limb variation of the solar \ion{Fe}{i} 7780\,\AA\ line, but they are unsuccessful for \ion{Fe}{i} 6151\,\AA. The metal-poor \avr\ model was found to be adequite for the strong \ion{Fe}{i} 5166\,\AA\ (\Eexc\ = 0) line, but inadequite in all other investigated cases. 
}

\maketitle{}

\section{Introduction}

Most stellar-parameters and chemical-abundance studies of cool stars still employ classical plane-parallel (1D) model atmospheres. How large are the abundance errors caused by ignoring hydrodynamical phenomena (3D effects) in the atmosphere of such stars?  
\citet{Collet2007} published (3D-1D) abundance corrections for some fictitious \ion{Fe}{i} and \ion{Fe}{ii} lines in a small grid of model atmospheres representing the atmosphere of cool giants of various metallicities. These data were updated by \citet{Hayek2011} by including Rayleigh scattering in the line-formation calculations. Significant changes were found only for the ultra-violet lines, with a wavelength of 3000~\AA. Therefore, for the lines lying in the visual spectral range, we cite hereafter the more extensive calculation of \citet{Collet2007}. 

   They showed that the 3D corrections are negative for the low-excitation (\Eexc\ $\le$ 2~eV) lines of \ion{Fe}{i} independent of their strength in the [Fe/H] $= -2$ and $-3$ models. The magnitude of the correction depends strongly on excitation energy of the lower level. For example, in the $\Teff$/$\logg$/[Fe/H] = 5035/2.2/$-2$ model, the (3D-1D) correction is slightly negative for the \Eexc\ = 4 and 5~eV lines, but it goes down to $-0.6$~dex for the \Eexc\ = 0 lines. The magnitude of correction increases towards lower metallicity. Classical abundance analysis of metal-poor (MP) stars should reveal an abundance difference of 0.5~dex between different excitation \ion{Fe}{i} lines if the theoretical predictions of \citet{Collet2007} are correct. 

  The theory predicts positive corrections for the \ion{Fe}{ii} lines.
For example, (3D-1D) = +0.2~dex for the \Eexc\ = 3~eV lines in the 5035/2.2/$-2$ model.
With 3D effects having opposite sign for \ion{Fe}{i} and \ion{Fe}{ii} lines, rather large  shifts to stellar surface gravities should result from spectroscopic analyses. This is worrisome given the overall decent agreement between spectroscopic analyses and stellar parameters from stellar-structure models. 

This study carefully checks the iron lines in the few MP stars with well determined stellar parameters using classical 1D model atmospheres and non-local thermodynamical equilibrium (NLTE) line formation. 
We are interested in uncovering any discrepancies between \ion{Fe}{i} and \ion{Fe}{ii} and between \ion{Fe}{i} lines of different excitation energy and to evaluate their magnitude if they exist.
The obtained results are presented in Sect.~\ref{sect:MP_1D}.

At present, there are no tools to perform full 3D-NLTE line formation calculations for atoms with a complicated term structure like those of \ion{Fe}{i} and \ion{Fe}{ii}. 
The full 3D line-formation calculations differ from their 1D counterparts due to different mean atmospheric structures and the existence of atmospheric inhomogeneities.
The effect of different mean atmospheric structures on emergent fluxes and derived chemical abundances can be evaluated using the spatially and temporally averaged atmospheric stratification denoted \avr. We investigate here how closely the \avr\ line formation for iron resembles that of the corresponding full 3D model. 
Section~\ref{sect:sun} compares the observed solar centre-to-limb variation of some \ion{Fe}{i} lines with the predictions from the \avr\ NLTE line-formation modelling. The LTE and NLTE calculations were also performed with the average 3D models representing the atmospheres of very metal-poor (VMP) cool giants (see Sect.~\ref{sect:3D}). Our conclusions are given in Sect.~\ref{sect:concl}.

\section{Method of calculations}

The statistical equilibrium of \ion{Fe}{i}-\ion{Fe}{ii} was calculated with a comprehensive model atom using the method of \citet{mash_fe} and the code {\sc DETAIL} \citep{detail}, where the opacity package was updated. 
Inelastic collisions with \ion{H}{i} atoms were taken into account employing the classical Drawinian rates scaled by a factor \kH\ = 0.1.
The line profile calculations were performed with the code {\sc SIU} \citep{Reetz}. 

For classical 1D model atmospheres, we used the {\sc MARCS} grid \citep{Gustafssonetal:2008} and the model 4600/1.6/$-2.5$ computed with the code {\sc MAFAGS-OS} \citep{Grupp2009}. The 3D model atmospheres were computed with the code ${\rm CO^5BOLD}$ \citep{Freytag2012}. The average \avr\ models were obtained by horizontally averaging each 3D snapshot over surfaces of equal (Rosseland) optical depth. For comparison, we also considered the 1D hydrostatic mixing-length model atmospheres which employ the
same micro-physics and radiative transfer scheme as ${\rm CO^5BOLD}$. They are referred to as ${\rm 1D_{LHD}}$ models. 

In this study, we determined absolute stellar iron abundances based on the linelist provided by \citet{mash_fe}. However, only the iron lines with accurate atomic data available were included in the analysis. For \ion{Fe}{i}, we selected the lines with $gf$-values measured by either the Oxford or Hannover group. The two different sets of $gf$-values were applied for \ion{Fe}{ii}. One is of \citet[][hereafter, MB09]{MB2009}, and another is of \citet[][hereafter, RUm]{RU1998}. The RUm $gf$-values were corrected by a factor +0.11~dex following the recommendation of \citet{Grevesse1999}. 

\section{1D-NLTE analysis of iron lines in some selected metal-poor stars}\label{sect:MP_1D}

\subsection{VMP cool giant HD~122563}

HD\,122563 was selected as one of the best observed halo stars, with well determined stellar parameters. Recent measurements of its angular diameter \citep{2012A&A...545A..17C} give $\Teff$ = 4600$\pm$41~K. In our earlier paper \citep{mash_fe}, we employed the same value, which was based on emprical colour calibrations,  to calculate the gravity $\logg$ = 1.60$\pm$0.07 from the {\sc HIPPARCOS} parallax. 

\begin{figure}
\flushleft 
  \resizebox{\hsize}{!}{\includegraphics[clip=true]{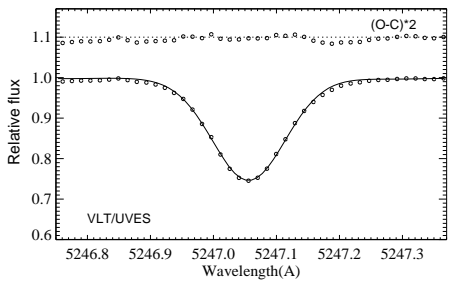}}
\caption{
\footnotesize
The best fit (continuous curve) of the \ion{Fe}{i} 5247~\AA\ line in HD\,122563 (open circles) obtained from the NLTE calculations with the {\sc MAFAGS-OS} model 4600/1.60/$-2.5$. The theoretical profile was convolved with a profile that combines instrumental broadening with a Gaussian profile of 3~\kms\ and broadening by macroturbulence with a radial-tangential profile of 4~\kms. The
differences between observed and calculated spectra (O - C) multiplied by a factor of 2 are presented in the upper part of figure (offset by +1.1). 
}\label{fe5247} 
\end{figure}

A high-quality spectrum of HD\,122563 ($R \simeq 80\,000$ and S/N $> 200$) was taken from the {\sc ESO UVESPOP} survey \citep{2003Msngr.114...10B}. The quality of line profile fits is illustrated in Fig.~\ref{fe5247} for \ion{Fe}{i} 5247~\AA. 

\begin{figure}
  \resizebox{\hsize}{!}{\includegraphics[clip=true]{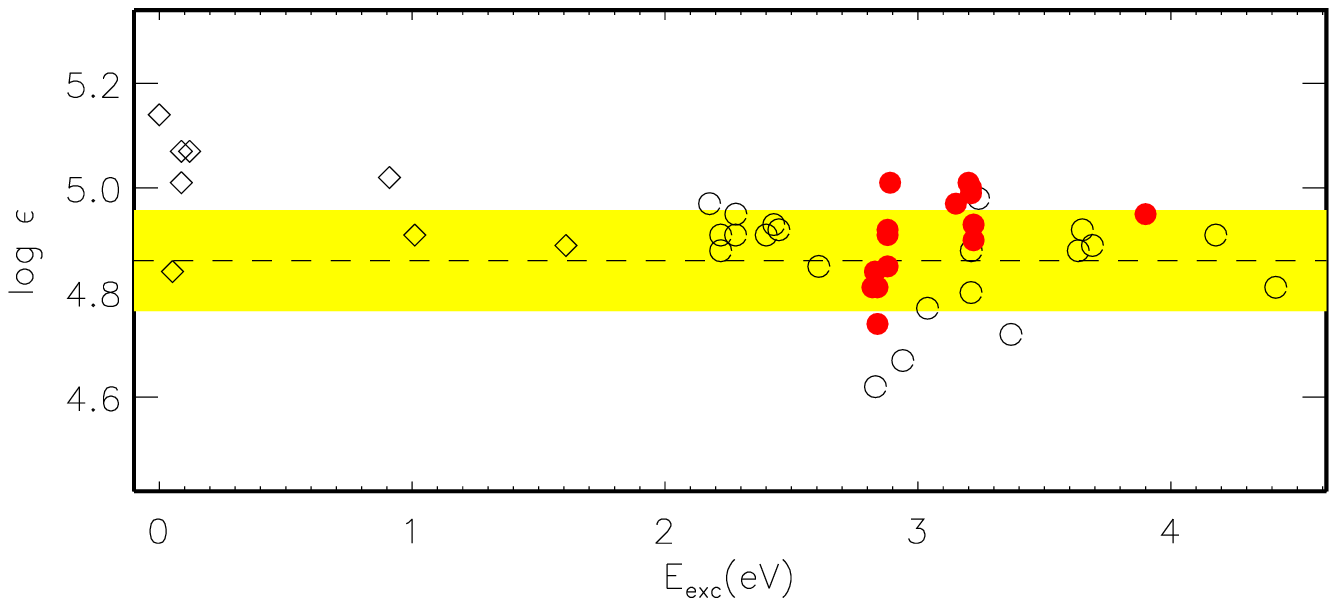}}
  \resizebox{\hsize}{!}{\includegraphics[clip=true]{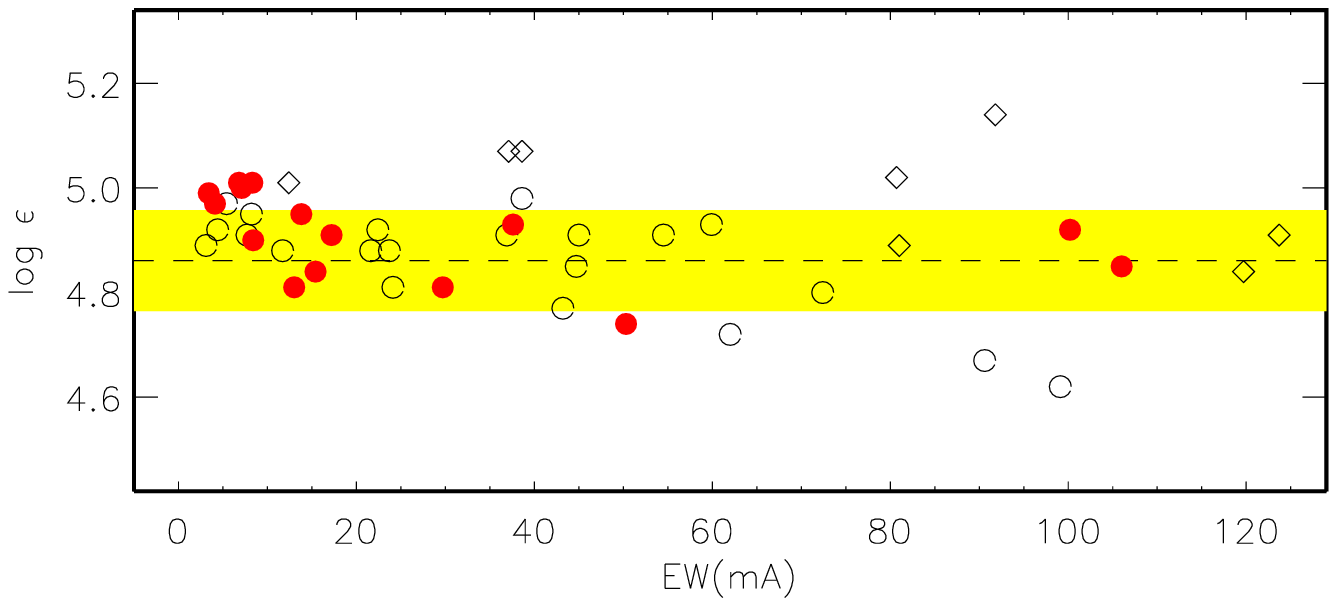}}
\caption{
\footnotesize
NLTE (\kH\ = 0.1) abundances derived from the \ion{Fe}{i} (open circles for \Eexc\ $\ge$ 2~eV and rombs for \Eexc\ $<$ 2~eV) and \ion{Fe}{ii} (filled circles) lines in HD\,122563 as a function of excitation energy of the lower level (top panel) and observed equivalent width (bottom panel). In each panel, the dashed line shows the mean iron abundance determined from the \ion{Fe}{i} \Eexc\ $\ge$ 2~eV lines.
}
\label{hd122563_abund}
\end{figure}

The NLTE abundances derived from individual lines in HD\,122563, in total, from 29 lines of \ion{Fe}{i} and 15 lines of \ion{Fe}{ii}, are presented in Fig.~\ref{hd122563_abund}. We adopted the microturbulence velocity $\xi_t$ = 1.95~\kms\ as determined by \citet{mash_fe} from \ion{Fe}{i} and \ion{Fe}{ii} lines. Interestingly, the use of different sets of $gf$-values leads to very similar mean abundances from the \ion{Fe}{ii} lines, i.e., $\eps{FeII}$(MB09) $= 4.92\pm0.08$ and $\eps{FeII}$(RUm) $= 4.91\pm0.09$. 
 When calculating the mean abundance from the \ion{Fe}{i} lines, we included only those with \Eexc\ $\ge$ 2~eV, which are predicted to be less affected by 3D effects. In NLTE (\kH\ = 0.1), we find consistent abundances between \ion{Fe}{i} and \ion{Fe}{ii}, with the difference (\ion{Fe}{i}($\ge$ 2~eV) - \ion{Fe}{ii})$_{\rm NLTE}$ = $-0.06$~dex, when using the MB09 $gf$-values. In contrast, the difference between the corresponding LTE abundances amounts to $-0.26$~dex. We note that the \ion{Fe}{i} lines arising from the ground term (\Eexc\ $<$ 0.2~eV) reveal, on average, a 0.2~dex higher abundance compared to that of the remaining \ion{Fe}{i} lines. 

 \citet{Collet2009} determined the LTE abundances from individual \ion{Fe}{i} and \ion{Fe}{ii} lines in HD\,122563 based on full 3D line-formation calculations. The stellar parameters employed were the same as in our study. For the \ion{Fe}{i} lines, \citet{Collet2009} found negative 3D abundance corrections, which depend strongly on the lower-level excitation potential. For example, (3D - 1D) $\simeq -0.3$ for \Eexc\ = 3~eV. In contrast, they obtained approximately 0.4~dex higher 3D than 1D abundances from the \ion{Fe}{ii} lines. \citet{Collet2009} argued that ``such trends may be attributed to the
neglected departures from LTE in the spectral line-formation calculations''.

When applying the 3D corrections of \citet{Collet2009} to our NLTE abundances, we obtain a dramatic discrepancy between \ion{Fe}{i} and \ion{Fe}{ii}, with (\ion{Fe}{i}($\ge$ 2~eV) - \ion{Fe}{ii})$_{\rm NLTE+3D}$ = $-0.75$~dex.

 As for the \ion{Fe}{i} lines arising from the ground state, 3D acts in right direction, but the predicted magnitude $-0.8$~dex of the 3D correction is markedly too large. 

\subsection{Subgiants and red giants of NGC~6397}

\begin{table*}
\caption{Iron abundances $\eps{}$ and abundance differences $\Delta$ = \ion{Fe}{i} -- \ion{Fe}{ii} for the stars in NGC\,6397}
\label{tab:ngc6397}
\begin{center}
\begin{tabular}{lcclllcccc}
\hline \hline \noalign{\smallskip}
\multicolumn{1}{c}{Star} & $\Teff$ & $\logg$ & \multicolumn{1}{c}{$\xi_t$} & \multicolumn{2}{c}{LTE} & & \multicolumn{3}{c}{NLTE} \\
\cline{5-6}
\cline{8-10}\noalign{\smallskip}
 & K & & \kms & \multicolumn{1}{c}{\ion{Fe}{i}} & \multicolumn{1}{c}{\ion{Fe}{ii}} & & \ion{Fe}{i} & \ion{Fe}{ii} & $\Delta$ \\
\noalign{\smallskip} \hline\noalign{\smallskip}
  & \multicolumn{9}{l}{IRFM 2010$^1$} \\
N5281  & 5978 & 3.73 & 1.75 & 5.31{\scriptsize{$\pm$0.07}} (15) & 5.34$^5${\scriptsize{$\pm$0.07}} (6) & & 5.52\scriptsize{$\pm$0.07} & 5.34$^5$\scriptsize{$\pm$0.07} & 0.18 \\
N8298  & 5978 & 3.73 & 1.75 & 5.31{\scriptsize{$\pm$0.06}} (15) & 5.36$^5${\scriptsize{$\pm$0.03}} (6) & & 5.52\scriptsize{$\pm$0.06} & 5.36$^5$\scriptsize{$\pm$0.03} & 0.16 \\
  & \multicolumn{9}{l}{IRFM 1999$^2$} \\
N5281  & 5816 & 3.69 & 1.75 & 5.19\scriptsize{$\pm$0.08} & 5.25$^4$\scriptsize{$\pm$0.09} & & 5.38\scriptsize{$\pm$0.08} & 5.25$^4$\scriptsize{$\pm$0.09} & 0.13 \\
N5281  & 5816 & 3.69 & 1.75 &                            & 5.31$^5$\scriptsize{$\pm$0.07} & &                            & 5.31$^5$\scriptsize{$\pm$0.07} & 0.07 \\
N8298  & 5816 & 3.69 & 1.75 & 5.19\scriptsize{$\pm$0.06} & 5.27$^4$\scriptsize{$\pm$0.09} & & 5.39\scriptsize{$\pm$0.06} & 5.27$^4$\scriptsize{$\pm$0.09} & 0.12 \\
N8298  & 5816 & 3.69 & 1.75 &                            & 5.33$^5$\scriptsize{$\pm$0.03} & &                            & 5.33$^5$\scriptsize{$\pm$0.03} & 0.06 \\
  & \multicolumn{9}{l}{IRFM 1996$^3$} \\
N11093 & 5100 & 2.51 & 1.7  & 5.09{\scriptsize{$\pm$0.10}} (15) & 5.16$^4${\scriptsize{$\pm$0.10}} (9) & & 5.27\scriptsize{$\pm$0.10} & 5.16$^4$\scriptsize{$\pm$0.10} & 0.11 \\
N11093 & 5100 & 2.51 & 1.7  &                            & 5.20$^5$\scriptsize{$\pm$0.06} & &                            & 5.20$^5$\scriptsize{$\pm$0.06} & 0.07 \\
N14592 & 5120 & 2.58 & 1.6  & 5.09{\scriptsize{$\pm$0.08}} (15) & 5.21$^4${\scriptsize{$\pm$0.09}} (9) & & 5.27\scriptsize{$\pm$0.09} & 5.21$^4$\scriptsize{$\pm$0.09} & 0.06 \\
N14592 & 5120 & 2.58 & 1.6  &                            & 5.25$^5$\scriptsize{$\pm$0.05} & &                            & 5.25$^5$\scriptsize{$\pm$0.05} & 0.02 \\
\noalign{\smallskip} \hline \noalign{\smallskip}
\multicolumn{10}{l}{Notes. Numbers in brackets indicate the total number of lines used in calculating the mean.} \\
\multicolumn{10}{l}{ \ $^1$ $\Teff$ is based on calibrations of \citet{Casagrande2010} for dwarfs and subgiants,} \\
\multicolumn{10}{l}{ \ $^2$ $\Teff$ is based on calibrations of \citet{Alonso1999} for dwarfs,} \\
\multicolumn{10}{l}{ \ $^3$ $\Teff$ is based on calibrations of \citet{Alonso1996} for giants,} \\
\multicolumn{10}{l}{ \ $^4$ based on $gf$-values of MB09;  \ $^5$ based on $gf$-values of RUm.} \\
  \end{tabular}
\end{center}
\end{table*}

Our analysis was extended to higher temperatures and higher gravities by inspecting the iron abundances of stars in the globular cluster (GC) NGC\,6397 on high-quality ($R \simeq 47\,000$, S/N $\simeq 100$) spectra taken with the FLAMES-UVES spectrograph at the Very Large Telescope \citep[see][for a detailed description of the observations]{Korn2007}. We selected the two stars N5281 and N8298 on the subgiant branch (SGB) and the two stars N11093 and N14592 on the red giant branch (RGB). 
Stellar parameters of these stars were discussed in detail by \citet{2012ApJ...753...48N}. Here, we employ the photometric effective temperatures deduced from two different empirical colour calibrations of \citet{Casagrande2010} and \citet{Alonso1999} for dwarfs and those from calibrations of \citet{Alonso1996} for giants. Following \citet{2012ApJ...753...48N}, we consider the SGB stars to belong to the dwarf calibration. It is worth noting that the calibrations of \citet{Casagrande2010} are also valid for subgiants.
The accurate determination of surface gravity of the GC stars takes advantage of their common distance that can be evaluated from the comparison of the observed and theoretical colour-magnitude diagrams. 
The gravity was computed from the usual relation between $\logg$ and $L$, $\Teff$, and mass of the observed stars. The used stellar parameters are indicated in Table~\ref{tab:ngc6397}. Microturbulence velocities were derived from \ion{Fe}{ii} lines by \citet{Korn2007}.

 The element abundances determined from individual iron lines turn out very similar for each pair of the SGB and RGB stars. 
As in HD\,122563, the \ion{Fe}{i} lines arising from the ground term give higher abundances compared to those from the other \ion{Fe}{i} lines, by 0.18~dex and 0.20~dex, on average, for the SGB and RGB stars, respectively. Therefore, mean \ion{Fe}{i} based abundances were computed ignoring the low-excitation (\Eexc\ $<$ 2~eV) lines of \ion{Fe}{i}. In contrast to HD\,122563, we find differences of 0.06~dex and 0.04~dex for the SGB and RGB stars, respectively, in mean abundance derived from the \ion{Fe}{ii} lines between applying $gf$-values from RUm and MB09. This is due to the smaller number of lines measured in the GC stars.
Table~\ref{tab:ngc6397} presents the obtained results.

 We find that the NLTE (\kH\ = 0.1) abundances determined from two ionization stages are consistent within the error bars in the RGB stars, when using the RUm $gf$-values, with (\ion{Fe}{i} - \ion{Fe}{ii})$_{\rm NLTE}$ = 0.07~dex and 0.02~dex for N11093 and N14592, respectively. In LTE, the corresponding differences amount to $-0.11$~dex and $-0.16$~dex. For the SGB stars, a discrepancy in NLTE abundances between \ion{Fe}{i} and \ion{Fe}{ii} does not exceed $1\sigma$ when employing the $\Teff$ scale of \citet{Alonso1999} and the RUm $gf$-values, but increases significantly for the hotter temperature $\Teff$ = 5978~K based on the \citet{Casagrande2010} calibrations.

When applying the 3D corrections presented by \citet{Korn2007}, we come to large discrepancy between \ion{Fe}{i} and \ion{Fe}{ii} not only in LTE, but also in NLTE, namely, (\ion{Fe}{i} - \ion{Fe}{ii})$_{\rm NLTE+3D}$ is equal $-0.2$~dex, on average, for the two RGB stars, $-0.2$~dex and $-0.1$~dex for the SGB stars on the $\Teff$ scale of \citet{Alonso1999} and \citet{Casagrande2010}, respectively.

 The above results may cast a shadow of doubt on 3D calculations. However, our interpretation is that one cannot simply add 1D-NLTE and 3D-LTE results when both NLTE and 3D effects are significant.
Exactly this is the case in metal-poor stars.

\section{Centre-to-limb variation of solar Fe~I lines}\label{sect:sun}

At present, one cannot perform full 3D and NLTE line-formation calculations for iron due to its complex atomic structure. In this section, we investigate whether NLTE calculations coupled to average \avr\ models can reproduce observations of the centre-to-limb variation of the solar \ion{Fe}{i} 7780\,\AA\ and 6151\,\AA\ lines. The measured equivalent widths ($EW$) were taken from \citet{Pereira2009}. We apply the ${\rm CO^5BOLD}$ solar 3D and average \avr\ models and also the MAFAGS-OS solar 1D model. When computing the theoretical $EW$, the iron abundance has been adjusted so that the models fit the observations at disk-centre. For the \avr\ and 1D model atmospheres, all the line profiles were computed with a microturbulence of $\xi_t$ = 0.9~\kms. 

\begin{figure*}
\resizebox{\hsize}{!}{\includegraphics[clip=true]{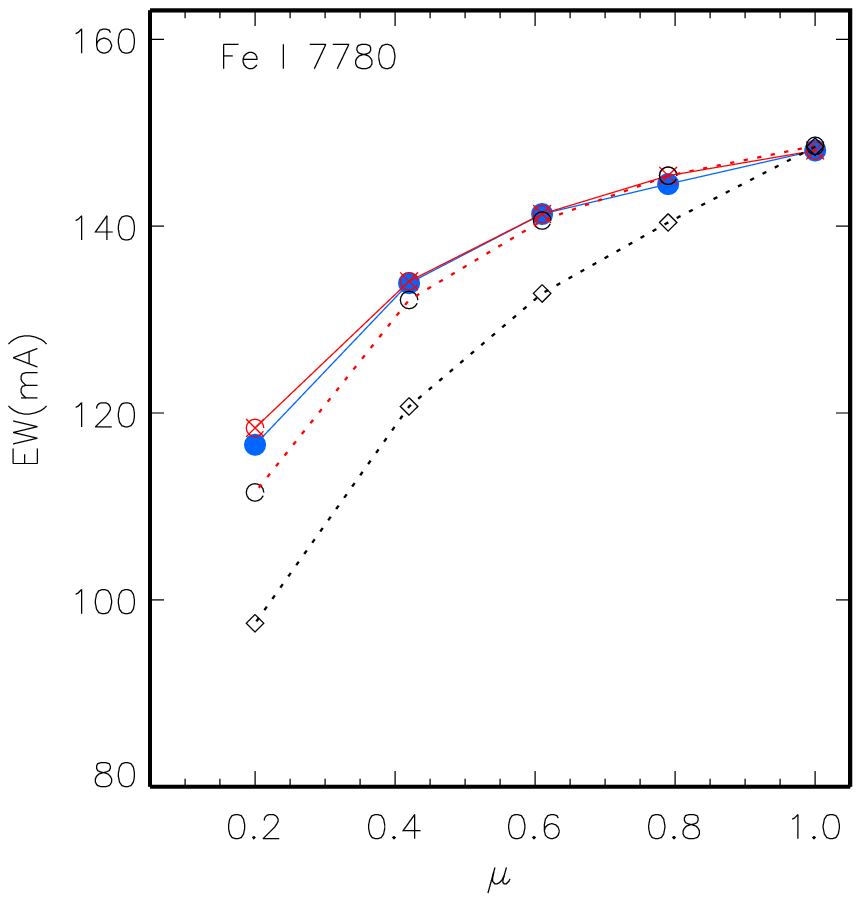}
\includegraphics[clip=true]{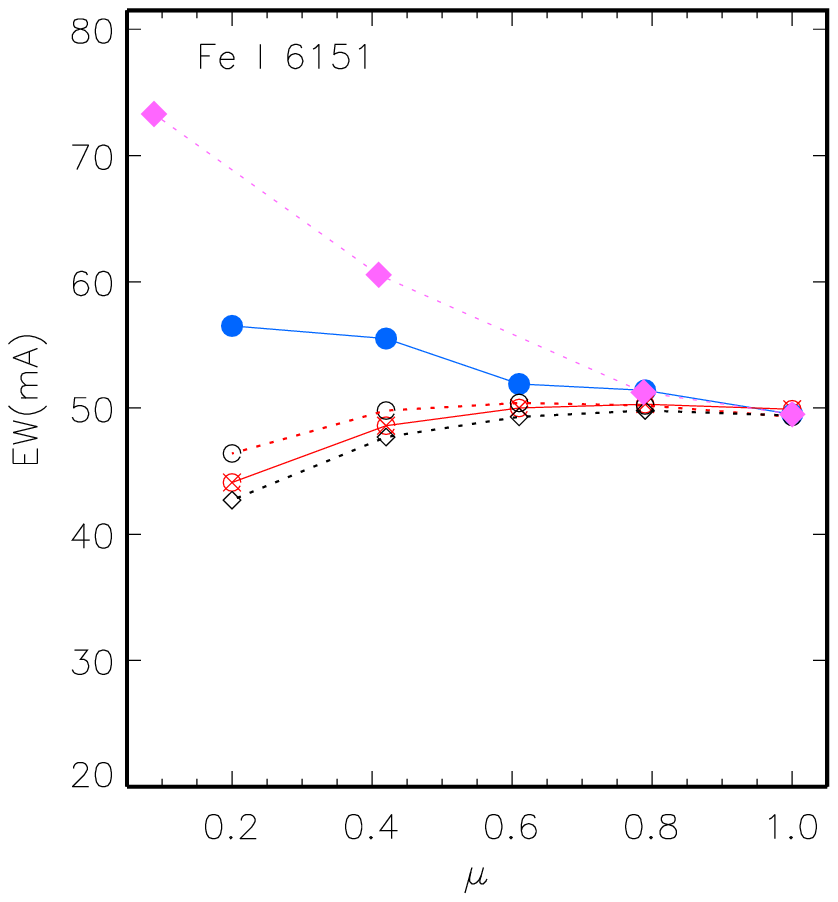}}
\caption{
\footnotesize
Solar observed (filled circles) and predicted centre-to-limb variations of line equivalent widths for \ion{Fe}{i} 7780\,\AA\ (left panel) and \ion{Fe}{i} 6151\,\AA\ (right panel). Filled rombs (only for \ion{Fe}{i} 6151\,\AA) correspond to full 3D-LTE line formation, crossed open circles to the NLTE calculations with the \avr\ model, open circles to LTE with \avr, and open rombs to LTE with the MAFAGS-OS model. 
}
\label{sun_int}
\end{figure*}

The obtained results are presented in Fig.\,\ref{sun_int}. It is evident that the classical model atmosphere fails to reproduce the observations. We do not show the NLTE results for the 1D model because NLTE leads to weakened \ion{Fe}{i} lines and larger discrepancy between the theory and observations. The NLTE calculations with the \avr\ model reproduce nearly perfectly the behavior of \ion{Fe}{i} 7780\,\AA, but the \avr\ model is unsuccessful in the case of \ion{Fe}{i} 6151\,\AA, both in LTE and NLTE. Full 3D-LTE line-formation calculations correctly predict the growth
of $EW$(\ion{Fe}{i} 6151\,\AA) toward the limb, albeit with overestimated theoretical $EW$s.
The latter can be due to neglecting the departures from LTE for \ion{Fe}{i}. 

These results illustrate the varying sensitivity of different \ion{Fe}{i} lines to 3D effects.
The \ion{Fe}{i} 7780\,\AA\ line seems to be mostly affected by the change in mean atmospheric structure of 3D compared with 1D model, while \ion{Fe}{i} 6151\,\AA\ is more sensitive to atmospheric inhomogeneity.

\section{Test calculations with metal-poor average \avr\ models}\label{sect:3D}

It would be useful to produce a list of lines that are more sensitive to the change in mean atmospheric structure than to atmospheric inhomogeneity. Such lines could be reliably employed in stellar abundance analyses with \avr\ models. In this section, we check three lines of \ion{Fe}{i} and \ion{Fe}{ii} 5018\,\AA\ line in the full 3D, average \avr, and reference 1D models of VMP cool giant with $\Teff$ = 4480~K, \logg\ = 2, and [M/H] = $-3$. The results are presented in Table\,\ref{tab:3D_corr} as the differences in abundance derived for a given line equivalent width between different models.

Despite the low iron abundance, all the selected lines are rather strong, with $EW$ from 70~m\AA\ to 95~m\AA. 
The full 3D-LTE line-formation calculations show that the effect of atmospheric inhomogeneity is weak only for \ion{Fe}{i} 5166\,\AA\ (the row 3D - \avr\ in Table\,\ref{tab:3D_corr}, part I). In contrast, \ion{Fe}{i} 5232\,\AA\ and \ion{Fe}{ii} 5018\,\AA\ are mostly affected by horizontal temperature/pressure fluctuations. To examine a dependence of the 3D effects on the line strength, the calculations were also performed with $gf$-values reduced by one order of magnitude (Table\,\ref{tab:3D_corr}, part II). 
The results differ dramatically: 3D effects are significantly reduced for \ion{Fe}{i} 5232\,\AA\ and \ion{Fe}{ii} 5018\,\AA\ and, in contrast, are enhanced
for the low-excitation lines, \ion{Fe}{i} 5166\,\AA\ and 
\ion{Fe}{i} 4994\,\AA. 
In the case labelled "weak lines", all lines are predominantly affected by atmospheric inhomogeneities
(the row 3D - \avr\ in Table\,\ref{tab:3D_corr}, part II). An exception is \ion{Fe}{i} 5232\,\AA, with its minor (3D - \avr) correction. 

\begin{table*}
\caption{Abundance differences (dex) between different 4480/2/$-3$ models}
\label{tab:3D_corr}
\begin{center}
\begin{tabular}{lrrrrr}
\hline \hline \noalign{\smallskip}
\multicolumn{2}{r}{Line:} & \ion{Fe}{i} 5166 & \ion{Fe}{i} 4994 & \ion{Fe}{i} 5232 & \ion{Fe}{ii} 5018 \\
\multicolumn{2}{r}{\Eexc (eV):} & 0.0    &    0.9   &    2.9  &    2.9 \\
\multicolumn{2}{r}{log $gf$:}   & $-4.20$ &  $-2.96$ & $-0.06$ & $-1.24$ \\
\noalign{\smallskip} \hline
 I. ``strong lines'' & $EW$(m\AA):   &   80   &   75     &    95   &     70 \\
 3D - ${\rm 1D_{LHD}}$ & {  LTE} &  $-0.22$ &   $-0.04$ &  +0.29 &     +0.50 \\
 3D - \avr &                   &  $-0.02$ &   +0.08   &  +0.35 &     +0.43 \\
 \avr$_{12}$ - ${\rm 1D_{LHD}}$ & {  NLTE} & 0.00 & +0.06 & +0.15 & +0.10 \\
                &  {  LTE}  & $-0.22$ & $-0.14$ & $-0.02$ & +0.10 \\
 \avr$_{12}$ - \avr & {  NLTE} & +0.03 & +0.04 & +0.05 & +0.02 \\
                 & {  LTE} & $-0.05$ & $ -0.04$ & $ -0.01$ & +0.02 \\
 \avr\ - ${\rm 1D_{LHD}}$ & {  NLTE} & $-0.07$ &  $-0.01$ &  +0.10 &     +0.09 \\
          &  {  LTE}  & $-0.16$ & $-0.09$ & $-0.01$ &     +0.08 \\
\noalign{\smallskip} \hline
 II. ``weak lines'' & $EW$(m\AA):   &   25   &   20     &    35   &     30 \\
 3D - ${\rm 1D_{LHD}}$ & {  LTE} & $-0.39$ &  $-0.23$ & +0.02 &  +0.20 \\
 3D - \avr &                   &  $-0.30$ &   $-0.20$ &  $-0.02$ & +0.13 \\
 \avr$_{12}$ - ${\rm 1D_{LHD}}$ & {  NLTE} &  +0.04 &   +0.08  & +0.13 & +0.10\\
                &  {  LTE} & $-0.15$ & $-0.08$  &  +0.04 & +0.10\\ 
 \avr$_{12}$ - \avr & {  NLTE} &  +0.03  &  +0.04  &     +0.05  & +0.02 \\
            &  {  LTE} & $-0.09$ &  $-0.06$  &    $-0.02$ & +0.02 \\ 
 \avr\ - ${\rm 1D_{LHD}}$ &  {  NLTE} &  +0.01 &   +0.04  &  +0.09  &+0.08 \\   
          &    {  LTE} &  $-0.06$ &   $-0.01$   &    +0.06 & +0.09 \\  
\noalign{\smallskip} \hline
  \end{tabular}
\end{center}
\end{table*}

We also performed LTE and NLTE calculations with various kinds of averaged 3D models. 
The full 3D model was decomposed into 12 groups depending on the emergent intensity. The fraction of surface area with given intensity defines the weight of the corresponding group.
By spatially and temporally averaging each group we obtained the 12 average group models. 
The theoretical emergent line fluxes from all the average group models were integrated using the group weights. The obtained results are hereafter referred to as \avr$_{12}$. 

\begin{figure}[]
  \resizebox{\hsize}{!}{\includegraphics[clip=true]{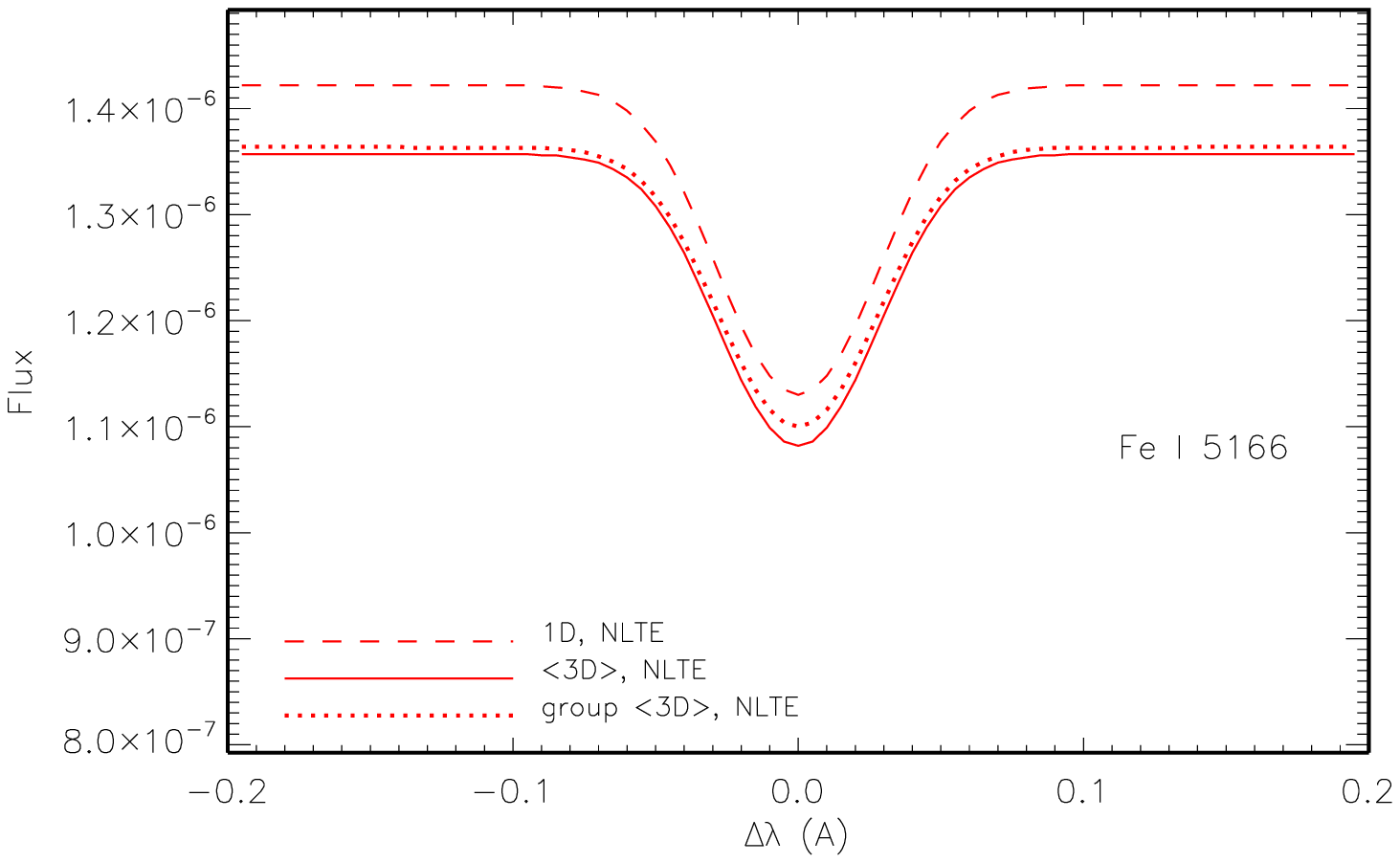}}
  \resizebox{\hsize}{!}{\includegraphics[clip=true]{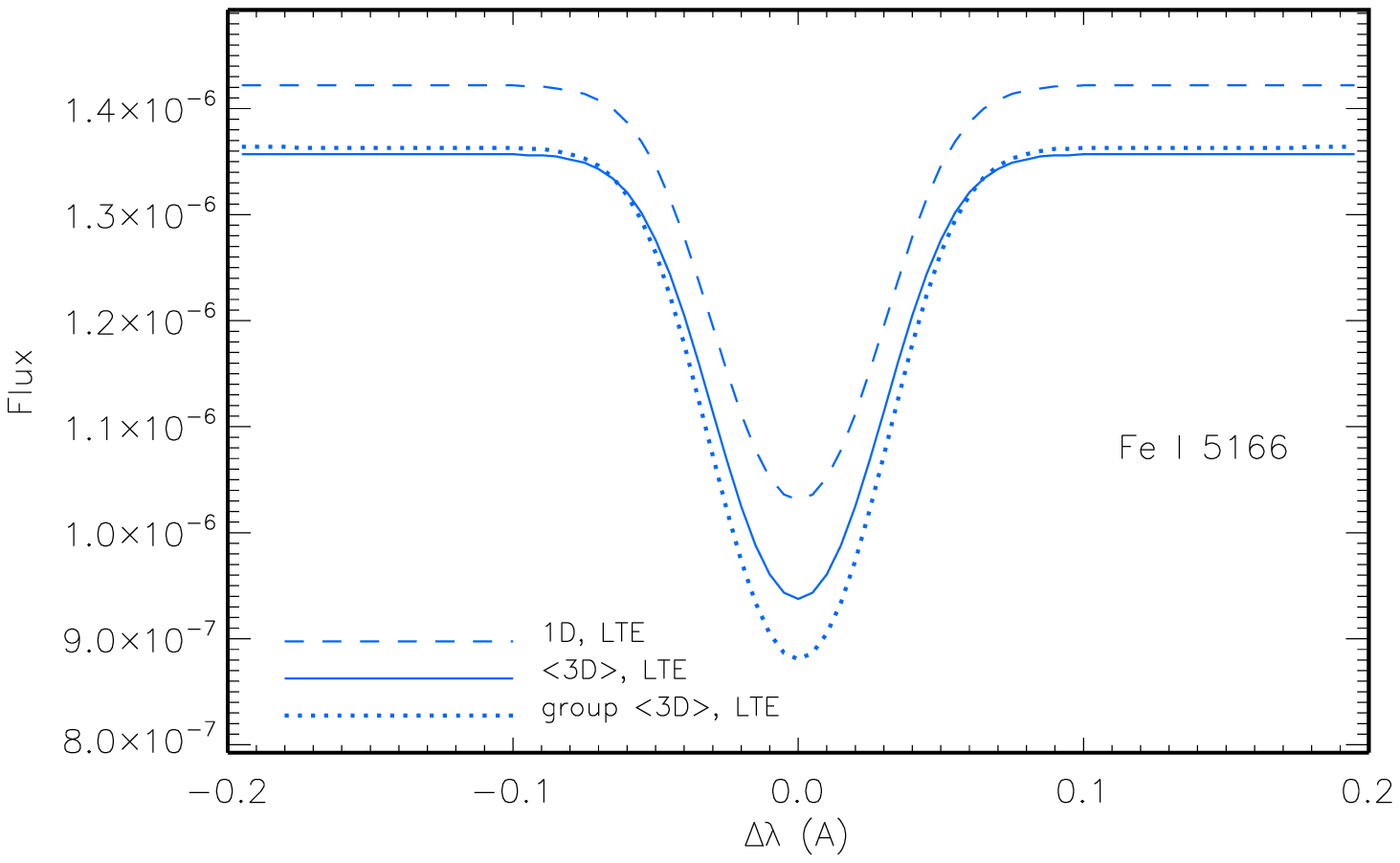}}
\caption{
\footnotesize
Theoretical NLTE (top panel) and LTE (bottom panel) fluxes (in erg\,cm$^{-2}$\,s$^{-1}$\,Hz$^{-1}$) for \ion{Fe}{i} 5166\,\AA\ from the \avr$_{12}$ (dotted curve), \avr\ (continuous curve), and ${\rm 1D_{LHD}}$ (dashed curve) models 4480/2/$-3$. All the computations were performed with log~$gf$ = $-5.2$ and $\xi_t$ = 1.8~\kms. 
}
\label{fig:fe5166_3d}
\end{figure}

In Fig.\,\ref{fig:fe5166_3d}, we compare the \ion{Fe}{i} 5166\,\AA\ line fluxes from different models in the ``weak line'' case. In LTE, the line is stronger in the \avr$_{12}$ model than in the \avr\ and 1D ones. In NLTE, the 3D effects are weakened and the \avr\ model approaches the \avr$_{12}$ one. In the ``strong line'' case, the difference (\avr$_{12}$ - ${\rm 1D_{LHD}}$) in LTE abundance is identical to (3D - ${\rm 1D_{LHD}}$) for this line.
The \avr$_{12}$ model seems to simulate the 3D model better than the \avr\ model for the lines affected mostly by the change in mean atmospheric structure. 

As for the remaining lines under the LTE assumption, both \avr$_{12}$ and \avr\ models are inadequate for \ion{Fe}{ii} 5018\,\AA\ and for \ion{Fe}{i} 5232\,\AA\ in the ``strong lines'' case. The 3D effects are, in part, reproduced by the \avr$_{12}$ and \avr\ models for \ion{Fe}{i} 4994\,\AA.

We find that the NLTE effects for each \ion{Fe}{i} line are stronger in the \avr$_{12}$ model than \avr\ and 1D ones. Only in the ``strong line'' case of \ion{Fe}{i} 5166\,\AA, where the \avr$_{12}$ model is working, the 3D effects are fully compensated by the NLTE effects, and no difference in NLTE abundances was obtained between \avr$_{12}$ and 1D models. 

We caution against simple adding the 1D-NLTE and 3D-LTE corrections to obtain the correction to the abundance derived from classical 1D-LTE analysis. For example, for the ``strong'' \ion{Fe}{i} 5166\,\AA\ line in the 4480/2/$-3$ model, we calculated (3D - 1D) = $-0.22$~dex under the LTE assumption. The NLTE correction computed with the 1D model amounts to $\Delta_{\rm NLTE}$ = $\eps{NLTE} - \eps{LTE}$ = 0.16~dex. Their adding results in a combined correction of $-0.06$~dex, while we see no difference in NLTE abundances between \avr$_{12}$ and 1D models. When applying $\Delta_{\rm NLTE}$ = 0.24~dex as computed with the \avr\ model, the combined NLTE and 3D correction is +0.02~dex. Therefore, the use of the average 3D model can be helpful, however, we stress, only when the 3D effects are mainly caused by the change in mean atmospheric structure.

\section{Conclusions}\label{sect:concl}

We summarize our findings as follows.
\begin{enumerate}
\item To within the error bars, consistent NLTE abundances were obtained from the two ionization stages, \ion{Fe}{i} and \ion{Fe}{ii}, for the stars with well determined stellar parameters HD\,122563 and red giants in NGC\,6397 using classical plane-parallel model atmospheres.
\item For HD\,122563 and the stars in NGC\,6397, the \ion{Fe}{i} lines arising from the ground term reveal approximately 0.2~dex higher abundances compared to those from other \ion{Fe}{i} lines. The obtained discrepancy between lines of different excitation lines cannot be explained by using the 3D corrections of \citet{Collet2007} and \citet{Hayek2011}. The 3D corrections predicted for the low-excitation \ion{Fe}{i} lines are markedly too large.
\item The use of the average \avr\ models is justified for particular \ion{Fe}{i} lines with the 3D effects mainly caused by the change in mean atmospheric structure of the 3D model compared to the 1D one. Our NLTE calculations reproduce well the centre-to-limb variation of the solar \ion{Fe}{i} 7780\,\AA\ line, but they are unsuccessful for  \ion{Fe}{i} 6151\,\AA. The metal-poor \avr\ model was found to be adequate for the ``strong'' \ion{Fe}{i} 5166\,\AA\ (\Eexc\ = 0) line, but inadequite for the lines sensitive to atmospheric inhomogeneities, i.e., \ion{Fe}{i} 5232\,\AA, \ion{Fe}{i} 4994\,\AA\, and \ion{Fe}{ii} 5018\,\AA. 
\item  For \ion{Fe}{i}, departures from LTE are expected to be stronger 
    in 3D models than in 1D ones.
\item Simply adding 1D-NLTE and 3D-LTE corrections to the abundances determined from classical LTE analysis does not provide 3D-NLTE results.
\end{enumerate}

Consistent 3D and NLTE line formation modelling is highly desirable for iron lines in metal-poor stars.

\begin{acknowledgements}
L.\,M. is supported by the DFG Sonderforschungsbereich (SFB) 881, ``The Milky Way System" and the Presidium RAS Programme ``Non-stationary phenomena in the Universe''. H.-G.\,L. is supported by the DFG SFB 881 (subproject A4). A.\,J.\,K. is supported by the Swedish National Space Board (SNSB) and the European Science Foundation (ESF). T.\,S. is supported by the Russian Federal Agency on Science and Innovation (grant 8394, 20.08.2012)
\end{acknowledgements}

\bibliography{cobold,ml_pb,nlte}

\begin{thebibliography}{20}
\expandafter\ifx\csname natexlab\endcsname\relax\def\natexlab#1{#1}\fi

\bibitem[{{Alonso} {et~al.}(1996){Alonso}, {Arribas}, \&
  {Martinez-Roger}}]{Alonso1996}
{Alonso}, A., {Arribas}, S., \& {Martinez-Roger}, C. 1996, \aap, 313, 873

\bibitem[{{Alonso} {et~al.}(1999){Alonso}, {Arribas}, \&
  {Mart{\'{\i}}nez-Roger}}]{Alonso1999}
{Alonso}, A., {Arribas}, S., \& {Mart{\'{\i}}nez-Roger}, C. 1999, \aaps, 140,
  261

\bibitem[{{Bagnulo} {et~al.}(2003){Bagnulo}, {Jehin}, {Ledoux}, {Cabanac},
  {Melo}, {Gilmozzi}, \& {ESO Paranal Science Operations
  Team}}]{2003Msngr.114...10B}
{Bagnulo}, S., {Jehin}, E., {Ledoux}, C., {et~al.} 2003, The Messenger, 114, 10

\bibitem[{{Butler} \& {Giddings}(1985)}]{detail}
{Butler}, K. \& {Giddings}, J. 1985, Newsletter on the analysis of astronomical
  spectra, No. 9, University of London

\bibitem[{{Casagrande} {et~al.}(2010){Casagrande}, {Ram{\'{\i}}rez},
  {Mel{\'e}ndez}, {Bessell}, \& {Asplund}}]{Casagrande2010}
{Casagrande}, L., {Ram{\'{\i}}rez}, I., {Mel{\'e}ndez}, J., {Bessell}, M., \&
  {Asplund}, M. 2010, \aap, 512, A54

\bibitem[{{Collet} {et~al.}(2007){Collet}, {Asplund}, \&
  {Trampedach}}]{Collet2007}
{Collet}, R., {Asplund}, M., \& {Trampedach}, R. 2007, \aap, 469, 687

\bibitem[{{Collet} {et~al.}(2009){Collet}, {Nordlund}, {Asplund}, {Hayek}, \&
  {Trampedach}}]{Collet2009}
{Collet}, R., {Nordlund}, {\AA}., {Asplund}, M., {Hayek}, W., \& {Trampedach},
  R. 2009, \memsai, 80, 719

\bibitem[{{Creevey} {et~al.}(2012){Creevey}, {Th{\'e}venin}, {Boyajian},
  {Kervella}, {Chiavassa}, {Bigot}, {M{\'e}rand}, {Heiter}, {Morel}, {Pichon},
  {Mc Alister}, {ten Brummelaar}, {Collet}, {van Belle}, {Coud{\'e} du
  Foresto}, {Farrington}, {Goldfinger}, {Sturmann}, {Sturmann}, \&
  {Turner}}]{2012A&A...545A..17C}
{Creevey}, O.~L., {Th{\'e}venin}, F., {Boyajian}, T.~S., {et~al.} 2012, \aap,
  545, A17

\bibitem[{{Freytag} {et~al.}(2012){Freytag}, {Steffen}, {Ludwig},
  {Wedemeyer-B{\"o}hm}, {Schaffenberger}, \& {Steiner}}]{Freytag2012}
{Freytag}, B., {Steffen}, M., {Ludwig}, H.-G., {et~al.} 2012, Journal of
  Computational Physics, 231, 919

\bibitem[{{Grevesse} \& {Sauval}(1999)}]{Grevesse1999}
{Grevesse}, N. \& {Sauval}, A.~J. 1999, \aap, 347, 348

\bibitem[{{Grupp} {et~al.}(2009){Grupp}, {Kurucz}, \& {Tan}}]{Grupp2009}
{Grupp}, F., {Kurucz}, R.~L., \& {Tan}, K. 2009, \aap, 503, 177

\bibitem[{Gustafsson {et~al.}(2008)Gustafsson, Edvardsson, Eriksson, Jorgensen,
  Nordlund, \& Plez}]{Gustafssonetal:2008}
Gustafsson, B., Edvardsson, B., Eriksson, K., {et~al.} 2008, A\&A, 486, 951

\bibitem[{{Hayek} {et~al.}(2011){Hayek}, {Asplund}, {Collet}, \&
  {Nordlund}}]{Hayek2011}
{Hayek}, W., {Asplund}, M., {Collet}, R., \& {Nordlund}, {\AA}. 2011, \aap,
  529, A158

\bibitem[{{Korn} {et~al.}(2007){Korn}, {Grundahl}, {Richard}, {Mashonkina},
  {Barklem}, {Collet}, {Gustafsson}, \& {Piskunov}}]{Korn2007}
{Korn}, A.~J., {Grundahl}, F., {Richard}, O., {et~al.} 2007, \apj, 671, 402

\bibitem[{{Mashonkina} {et~al.}(2011){Mashonkina}, {Gehren}, {Shi}, {Korn}, \&
  {Grupp}}]{mash_fe}
{Mashonkina}, L., {Gehren}, T., {Shi}, J.-R., {Korn}, A.~J., \& {Grupp}, F.
  2011, \aap, 528, A87

\bibitem[{{Mel{\'e}ndez} \& {Barbuy}(2009)}]{MB2009}
{Mel{\'e}ndez}, J. \& {Barbuy}, B. 2009, \aap, 497, 611

\bibitem[{{Nordlander} {et~al.}(2012){Nordlander}, {Korn}, {Richard}, \&
  {Lind}}]{2012ApJ...753...48N}
{Nordlander}, T., {Korn}, A.~J., {Richard}, O., \& {Lind}, K. 2012, \apj, 753,
  48

\bibitem[{{Pereira} {et~al.}(2009){Pereira}, {Asplund}, \&
  {Kiselman}}]{Pereira2009}
{Pereira}, T.~M.~D., {Asplund}, M., \& {Kiselman}, D. 2009, \aap, 508, 1403

\bibitem[{{Raassen} \& {Uylings}(1998)}]{RU1998}
{Raassen}, A.~J.~J. \& {Uylings}, P.~H.~M. 1998, \aap, 340, 300

\bibitem[{Reetz(1991)}]{Reetz}
Reetz, J.~K. 1991, Diploma Thesis (Universit\"at M\"unchen)

\end{thebibliography}
\bibliographystyle{aa}

\end{document}